\documentclass[%
aip,
amsmath,amssymb,
preprint
]{revtex4-2}
\DeclareUnicodeCharacter{2212}{\textminus}
\usepackage{subcaption}
\usepackage{placeins}
\usepackage{graphicx}
\usepackage{dcolumn}
\usepackage{bm}

\usepackage[utf8]{inputenc}
\usepackage[T1]{fontenc}
\usepackage{mathptmx}
\usepackage{etoolbox}
\usepackage[english]{babel}
\makeatletter

\def\bbl@set@language#1{%
	\edef\languagename{%
		\ifnum\escapechar=\expandafter`\string#1\@empty
		\else\string#1\@empty\fi}%
	\@ifundefined{babel@language@alias@\languagename}{}{%
		\edef\languagename{\@nameuse{babel@language@alias@\languagename}}%
	}%
	\select@language{\languagename}%
	\expandafter\ifx\csname date\languagename\endcsname\relax\else
	\if@filesw
	\protected@write\@auxout{}{\string\select@language{\languagename}}%
	\bbl@for\bbl@tempa\BabelContentsFiles{%
		\addtocontents{\bbl@tempa}{\xstring\select@language{\languagename}}}%
	\bbl@usehooks{write}{}%
	\fi
	\fi}
\newcommand{\DeclareLanguageAlias}[2]{%
	\global\@namedef{babel@language@alias@#1}{#2}%
}
\def\@email#1#2{%
	\endgroup
	\patchcmd{\titleblock@produce}
	{\frontmatter@RRAPformat}
	{\frontmatter@RRAPformat{\produce@RRAP{*#1\href{mailto:#2}{#2}}}\frontmatter@RRAPformat}
	{}{}
}%
\makeatother
\DeclareLanguageAlias{en}{english}
\begin{document}
	
	\preprint{AIP/123-QED}
	
	\title{Poly- and single-crystalline diamond nitrogen-induced TLS losses estimation with superconducting lumped elements micro-resonators.}
	
	\author{F. Mazzocchi}
	\affiliation{Institute of Advanced Materials - Advanced Materials Physics (IAM-AWP), Karlsruhe Institute Of Technology.}
	\email{francesco.mazzocchi@kit.edu}
	
	\author{M. Neidig}%
	\affiliation{ 
		Institute Of Micro- and Nanoelectronic Systems (IMS),  Karlsruhe Institute Of Technology.
	}%
	
	\author{H. Yamada}%
	\affiliation{ 
		National Institute of Advanced Industrial Science and Technology (AIST),  Osaka.
	}
	
	\author{S. Kempf}
	\affiliation{ 
		Institute Of Micro- and Nanoelectronic Systems (IMS), Karlsruhe Institute Of Technology.
	}%

	\author{D. Strauss}
	\affiliation{Institute of Advanced Materials - Advanced Materials Physics (IAM-AWP), Karlsruhe Institute Of Technology.}

	\author{T. Scherer}
	\affiliation{Institute of Advanced Materials - Advanced Materials Physics (IAM-AWP), Karlsruhe Institute Of Technology.}
	
	\date{\today}
	
	\begin{abstract}
		Research on diamond has intensified due to its exceptional thermal, optical, and mechanical properties, making it a key material in quantum technologies and high-power applications. Diamonds with engineered nitrogen-vacancy (NV) centers represent a very sensitive platform for quantum sensing, while high-optical quality diamond windows represent a fundamental safety component inside Electron Cyclotron Resonance Heating (ECRH) systems in nuclear fusion reactors. A major challenge is the development of ultra-low-loss, high-optical-quality single-crystal diamond substrates to meet growing demands for quantum coherence and power handling. Traditionally, dielectric losses ($\tan \delta$) in diamonds  are evaluated using Fabry-Perot microwave resonators, in which the resonance quality factors Q of the cavity with and without the sample are compared. These devices are limited to resolutions around 10$^{-5}$ by the need to keep the resonator´s dimensions within a reasonable range. In contrast, superconducting thin-film micro-strip resonators, with Q factors exceeding 10$^6$, are stated to provide higher sensitivity for assessing ultra-low-loss materials. This study examines four diamond samples grown through different processes, analyzing their dielectric losses at extreme low temperatures (sub-Kelvin) within the Two-Level System (TLS) framework. Complementary Raman spectroscopy measurements allowed us not only to associate higher nitrogen content with increased losses, but also to investigate how the different growth process influence the way these defects are incorporated in the crystal lattice.
	\end{abstract}
	
	\maketitle
	
	\begin{quotation}
		High-Q superconducting resonators offer a promising platform for determining the dielectric losses ($\tan \delta$) of ultra-low-loss materials, surpassing the resolution limits of state-of-the-art Fabry-Perot resonators. Materials such as poly- and single-crystalline diamond play a critical role in applications ranging from nuclear fusion to quantum sensing, where exceptionally high optical quality is essential. In this work, we present the characterization of four diamond samples grown using different techniques, leveraging superconducting micro-strip resonators to assess their dielectric properties. The measurements are then interpreted within the Two-Level System (TLS) theory to evaluate this particular loss channel.
	\end{quotation}
	
	\section{\label{sec:intro}Introduction\protect\\} 
	Electron Cyclotron Resonance Heating (ECRH) systems are crucial in modern fusion reactors, providing efficient heating and plasma control through high-power microwave radiation \cite{Dammertz2004,Laqua2021}. A key component is the diamond window \cite{Brandon2001}, which serves as a physical barrier between the reactor- and the gyrotron-side of the ECRH system. Poly-crystalline diamond windows are commonly used in ECRH systems due to their excellent microwave transparency and high thermal conductivity. These windows must meet stringent requirements, including low losses, to minimize microwave power attenuation \cite{Schreck2015,Schreck2018}. The loss tangent, which measures microwave energy absorption, is typically determined using Fabry-Perot resonators \cite{Heidinger2002,Scherer2011}. While poly-crystalline diamond windows meet current ECRH performance needs \cite{Aiello2019}, single-crystal diamond windows \cite{schreck2014} offer lower loss tangents, making them the natural successor of poly-crystalline discs for future reactors with higher power demands\cite{Jelonnek2017,Avramidis2019,Denisov2022}. However, their ultra-low loss tangent exceeds the measurement capabilities of traditional Fabry-Perot resonators. Nitrogen-doped diamond, which incorporates nitrogen-vacancy (NV) color centers, is also of great interest for quantum sensing applications, as NV pairs create quantum systems with long coherence times \cite{burgler2023all} that depend also on the substrate optical losses \cite{Kim2015}. Superconducting resonators, inspired by Kinetic Inductance Detector (KID) technology \cite{Roesch2014, Mazzocchi2021,Mazzocchi2023}, offer the sensitivity needed to measure extremely low losses in single-crystal diamond windows given their extremely high Q-factors \cite{Heidari2019,Baselmans2007}. Additionally, since they are already implemented in NV-based qubit systems \cite{Vallabhapurapu2021} for microwave power delivery and spin control, they could also serve as in-situ probes for real-time measurement of dielectric losses and noise levels within the qubit system. 
	\section*{Loss Tangent and TLS}\label{tang}
\begin{figure*}
		\includegraphics[width=\textwidth]{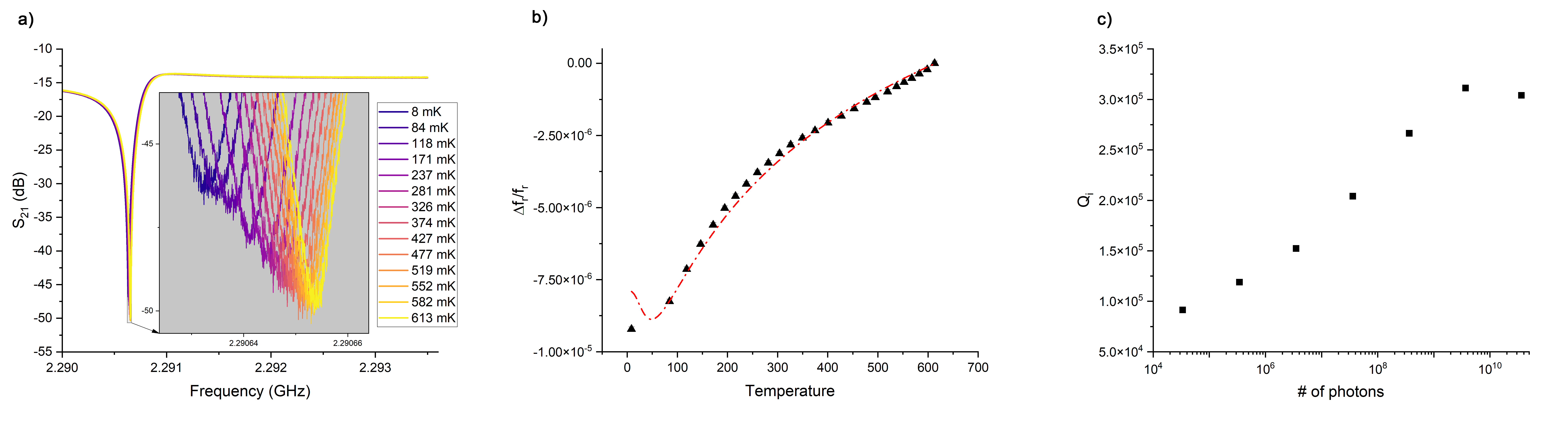} 
		\caption{a) example of $f_r$ vs T measurement performed in the ultra-low temperature range on one of the samples. Warmer colors indicate higher acquisistion temperatures b) the fractional resonance frequency shift associated with the measurement set shown in Fig. 1a. The dotted red line identifies the fitting curve that is used to extract the value of the losses c) resonator internal quality factor vs number of photons injected, for on-chip powers ranging from -45  to -115 dBm (10 dBm steps) and a calculated single photon energy around -150 dBm. Visible is the onset of $Q_i$ saturation above 3.5E9 ($\approx$ -55 dBm) photons present in the resonator}
		\label{pic1}
	\end{figure*}

	The \textit{loss tangent} $\tan \delta$ is a parameter used to describe the dielectric losses in a material when subjected to an alternating electric field. According to the Debye model \cite{Debye1954}, it is defined as the ratio of the imaginary part of the complex permittivity, $\epsilon = \epsilon' - j\epsilon''$, to the real part:
	\begin{equation}
		\tan \delta = \frac{\epsilon''}{\epsilon'}.
	\end{equation}
	Here, $\epsilon'$ represents the ability of the material to store electrical energy, while $\epsilon''$ accounts for the energy dissipated. 
	One of the methods to experimentally determine the loss tangent of a dielectric substrate involves Fabry-Perot microwave open resonators in various configurations. In such techniques, the Q-factor and the resonant frequency of the cavity with and without the sample are compared, since the $\tan \delta$ relates directly to the total quality factor of a resonator through the following relationship
	\begin{equation}\label{tan}
		\tan \delta = \frac{1}{Q}.
	\end{equation} 
	The resolution power of these devices is limited by requirements for reasonable dimensions, with table-top double spherical systems \cite{Scherer2011} capable of reaching precision in the order of $10^{-5}$. Single crystal diamond substrates in development to improve upon this parameter will present losses too low to be resolved by the aforementioned devices. A suitably higher resolution measurement technique is therefore necessary for their characterization. Superconducting micro-strip resonators have shown intrinsic Q factors routinely above $10^6$ with record values approaching the $10^7$ range \cite{Heidari2019} and represent a valid option to supersede open cavity resonators for the study of ultra-low loss dielectric materials. They consist of a substrate, either a material with well-known characteristics or the sample itself, over which a superconducting thin film is deposited and patterned. The resulting circuit can take the form of several different distributed and also lumped elements configurations. The total (or "loaded") quality factor $Q_L$ is given by a combination of the intrinsic $Q_i$ associated with the losses of the resonator itself and the coupling $Q_C$ associated  with the coupling between the circuit and the feed-line used to excite it \cite{Francesco2022}:
	\begin{equation}
		\frac{1}{Q_L} = \frac{1}{Q_i} + \frac{1}{Q_C} 
	\end{equation}
	The intrinsic $Q_i$ encompasses contributions from multiple sources of loss, including conductor losses ($Q_c$), radiation losses ($Q_r$), and dielectric losses ($Q_d$):
	
	\begin{equation}
		\frac{1}{Q_i} = \frac{1}{Q_c} + \frac{1}{Q_r} + \frac{1}{Q_d}.
	\end{equation}
	Radiative losses can be limited through design of the device (cfr. section \ref{des}), while conduction and dielectric losses can dominate over each other depending on the measurement conditions. One way to discern in which regime one operates is to observe the behavior of the resonance frequency vs temperature variation. Generally, in the low temperature regime (3.5 - 9 K in our case) the superconductor behaves "classically", with lower temperatures associated to a resonance frequency blue-shift. This is due to the variation of the superconductor´s kinetic inductance $L_k$ \cite{Tinkham2004} as the quasi-particles generation-recombination dynamic equilibrium is shifted by the changing temperature. It´s a clear indication that the temperatures are still too high and the conductor losses dominate over the dielectric ones. On the other side, measurements performed in the ultra-low range (sub-Kelvin) outline an opposite trend, with frequency blue-shifts associated to higher temperatures (Fig.\ref{pic1}a). This behavior has been repeatedly associated with two levels systems (TLS) dynamics \cite{Sueno2022, Burnett2014}. Defects in the crystal lattice can create localized states in the material's energy spectrum, leading to increased TLS activity and associated energy dissipation. At higher temperatures, more TLS are promoted to higher energy states, reducing the number of available low-energy states that could otherwise absorb microwave energy. As a result, microwave-induced losses decrease, leading to an overall enhancement in resonator performance. Under these premises, we can introduce a temperature-dependent variation of the permittivity given by \cite{Gao2008}:
	\begin{equation}\label{TLS}
		\frac{\Delta \varepsilon}{\varepsilon} = - \frac{2 \tan \delta^0_{TLS}}{\pi} \left[ \Re \Psi \left( \frac{1}{2} + \frac{1}{2 \pi i} \frac{\hbar \omega}{k_B T} \right) - \log \frac{\hbar \omega}{k_B T} \right]
	\end{equation}
	with $\omega$ represents the frequency, $\Psi$ is the complex digamma function and $\tan \delta^0_{TLS} = \pi P d^2/3$ is the loss tangent at zero temperature and power for a TLS ensemble with average electrical dipole moment $d$ and $P$ density of states. Assuming an isotropic distribution of the TLSs, we can put the variation of the permittivity directly in relation with the fractional resonance shift (Fig. \ref{pic1}b) observed in the measurements:
	\begin{equation}
		\frac{\Delta f_r}{f_r} = - \frac{q}{2} \frac{\Delta \varepsilon}{\varepsilon}
	\end{equation}
	where q represents the filling factor that accounts for the distribution of the electric field between the substrate and the surrounding air. 
	The filling factor can be directly calculated if we consider that micro-strip resonators support quasi-transverse electromagnetic (quasi-TEM) modes, enabling the use of the quasi-static approximation. Under this model, we can define an effective permittivity ($\varepsilon_{\text{eff}}$) that accounts for the electromagnetic field distribution between the substrate and air and it´s a function of the resonator geometry. It reflects a weighted average between the substrate's relative permittivity ($\varepsilon_r$) and air's permittivity ($\varepsilon_0 = 1$), and is given by \cite{Hammerstad1980, Schneider1969}:
	\begin{equation}\label{eeff}
		\varepsilon_{\text{eff}} = \frac{\varepsilon_r + 1}{2}+\frac{\varepsilon_r - 1} \cdot \frac{1}{\sqrt{10\frac{h}{w}}}
	\end{equation}
	where h and w represent substrate thickness and micro-strip width, respectively. The filling factor can be at this point estimated with the following equation \cite{Gupta1996}:
	\begin{equation}
		q = \frac{\varepsilon_{\text{eff}}-1}{\varepsilon_r - 1}
	\end{equation}

	This way, by measuring the frequency shift we are able to gather information about the substrate quality in terms of TLS activity, with the only requirement being to operate at sub-Kelvin temperatures to minimize conductor losses. TLS occupancy will eventually start saturating as the injected power is increased, up to the point where the losses will start rising again as a consequence of increased $L_k$ (Fig. \ref{pic1}c).
	
	\section{Device design and setup}\label{des}
	\begin{figure*}[b] 
		\includegraphics[width=\textwidth]{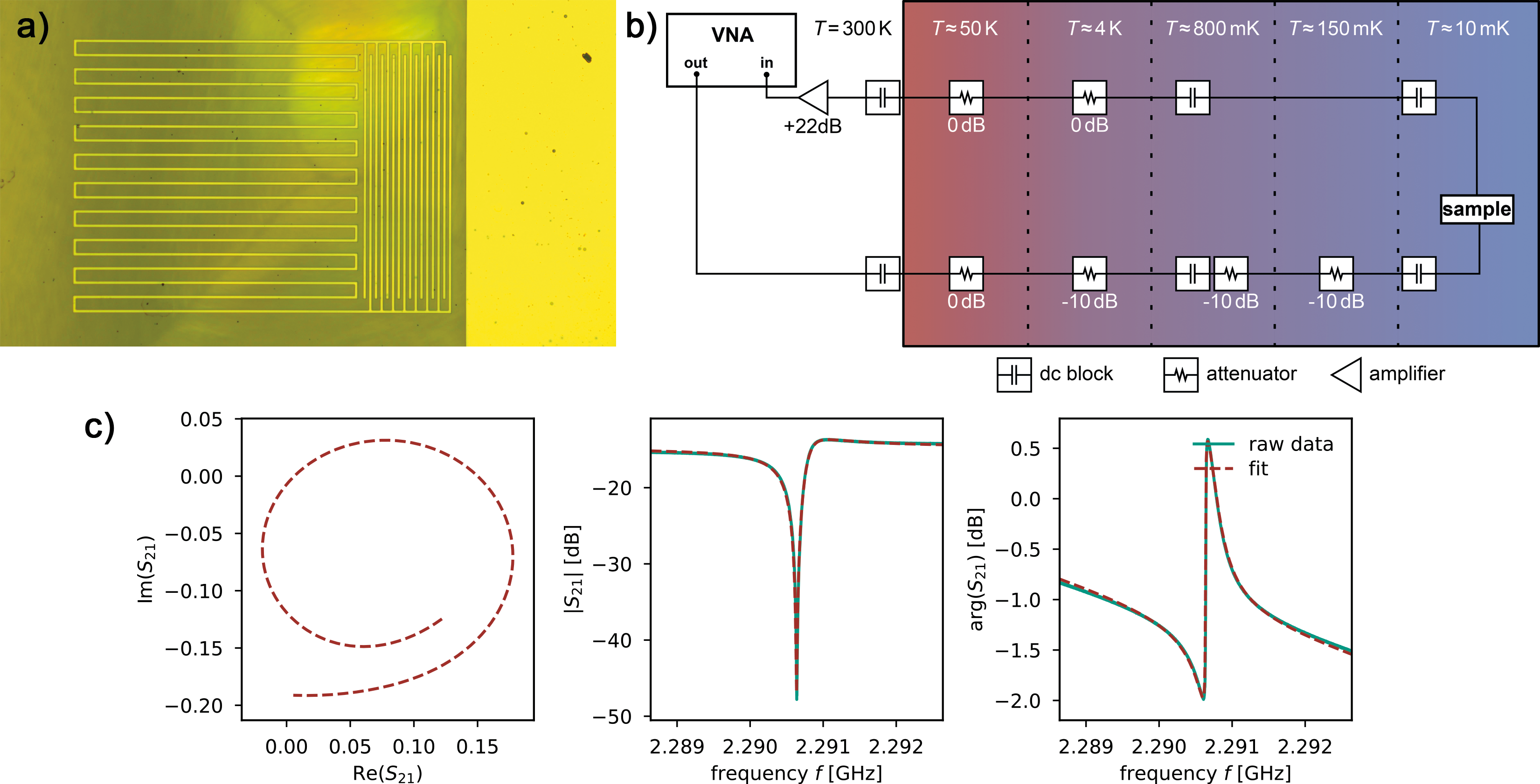}
		\caption{a) optical microscope image of one of the lumped elements resonators used in this work. Visible are the meandering inductor, the IDC and a section of the feed-line  b) schematic representation of the acquisition setups used in the dilution refrigerator  c) An example of the circle-fits correction routines employed in this work}
		\label{pic2}
	\end{figure*}
	After an initial simulation study with the SONNET software suite \cite{Rautio2021, Ney1985, Gibson2021} in which several different configurations were analyzed, a Lumped Elements (Fig.\ref{pic2}a) kind of resonator was in the end selected in virtue of its higher Q-factors and tuning flexibility. It is also well known that in LE resonators the current is mostly concentrated (and uniformly distributed \cite{Francesco2022}) in the meandering inductor, with almost none flowing in the inter-digital capacitor (IDC). This, apart from enabling detectors fashioned from these devices to operate without an antenna, has the distinct advantage of greatly limiting the radiation losses, that therefore will not be considered. The 300 nm thick niobium film was deposited on the growth side of all samples via DC magnetron sputtering, patterned with a direct laser-scribing machine and etched with a reactive ion etching chamber working with a SF6/O2 gas mixture. Niobium is the elemental superconductor with the highest critical temperature, therefore it represents a valid candidate for our application. The micro-strip composing the IDC and the inductor has a uniform width of 10 $\mu$m, which translates in a calculated effective permittivity (Eq. \ref{eeff}) equal to approximately 3.81, considering a substrate with $\varepsilon_r$ = 5.67. This value is in excellent agreement with the one calculated via the simulations, equal to 3.6. 
	The RF characterization of the patterned samples has been carried on in two different temperature ranges: low (approx 3.5 to 9 K) and ultra-low (10-700 mK) in a Pulse Tube Cooler (PTC) and Dilution Refrigerator (DR) respectively. Given the arguments exposed in section \ref{sec:intro}, only the sub-Kelvin measurement will be shown. Fig. \ref{pic2}b shows the acquisition setup scheme used in the DR. The sample is mounted to the mixing chamber stage of the cryostat at millikelvin temperatures. The transmission parameter $S_{21}$ is measured using a vector network analyzer (VNA), which is connected to the sample through the cryogenic microwave readout chain. Installed dc blocks prevent undesired ground loops and provide thermal isolation between the temperature stages. The VNA output signal is attenuated at low temperatures to optimize the signal-to-noise ratio at the reduced power levels used for on-chip readout. At room temperature, the microwave signal is amplified using a low noise amplifier. Additional $0\,\mathrm{dB}$ attenuators improve the thermalization of the coaxial cable’s inner conductor and minimize heat load at lower temperatures.
	From the measured data, key characteristics of the microwave resonator---such as resonance frequency, loaded quality factor, and internal quality factor---can be extracted using a numerical fitting algorithm \cite{Probst2015}. The complex transmission scattering parameter, $S_{21}$, around the resonance frequency  $f_r$, is modeled as a function of frequency $f$ as follows:
	
	\begin{equation}\label{circlefit}
		S_{21}(f) = a e^{i\alpha} e^{-2 \pi i f \tau} \left[ 1 - \frac{(Q_L/|Q_C|)e^{i\phi}}{1+2 i Q_L (f/f_r-1)} \right]
	\end{equation}
	
	This model takes into account various system-related effects, including electronic delay $\tau$, attenuation $a$, phase shift $\alpha$, and any potential impedance mismatch $\phi$ between the transmission line and the microwave resonator. An exemplary fit of the model to the acquired data is illustrated in Fig. \ref{pic2}c.

	\section{Experimental}\label{res}
	\subsection{The samples}
	Four different diamond samples, each 350 $\mu$m thick and approximately 5x5mm, were investigated in this work. Our reference consisted in a poly-cristalline diamond (PCD) sample produced by Diamond Materials company in Freiburg \cite{diam2024}, Germany. These substrate are the current golden standard for high-power microwave applications and their specifications include a ambient temperature $ \tan \delta \approx 10^{-5}$ for radiation in the 100-200 GHz range. They are grown with a Chemical Vapor Deposition (CVD) process, and the deposition rate can be boosted by adding nitrogen into the reaction chamber \cite{MuellerSebert1996, Chayahara2004, Achard2007}, with the trade off being the formation of smaller grains and, therefore, increased losses. Our sample is of the highest “optical grade” quality, with an average grain size on the growth side equal to 50 $\mu$m \cite{Aiello2024}. The other three samples where all single-crystal diamonds, grown following different processes. The first of these is a Hetero-epitaxial Single-crystal Diamond grown on iridium (IrSCD). In this process, iridium's lattice structure and chemical properties support the nucleation and growth of single-crystal diamond, while also having low tendency and affinity to form stable carbide compounds that would lead to defects and lower quality. Therefore, as carbon atoms accumulate on the substrate, they segregate to the surface, forming the basis for diamond layer growth. The controlled nature of this segregation minimizes unwanted inclusions and defects in the final crystal \cite{Verstraete2005, Schreck2001, Schreck2017}. The second single-crystal specimen (marked as "Clone") was a lift-off clone produced with an ion-implantation–assisted CVD growth process on a type IIa HPHT seed \cite{Mokuno2014}, followed by restorative overgrowth. The last single crystal sample was also produced with the clone technique and then laser-cut in half.  After re-alignment of the two halves, the interfaces were re-joined through CVD and underwent an additional lift-off to obtain a free-standing wafer (marked as J-Clone) \cite{Yamada2012, Yamada2013, Yamada2015}. Repeating such growth and joining processes enables the production of inches size single-crystal wafers while preserving dislocation density and optical quality. Both clone substrates were provided by the AIST in Osaka, Japan.
	
	\subsection{Results and Discussion}
	
\begin{figure*}[h!]
		\includegraphics[width=\textwidth]{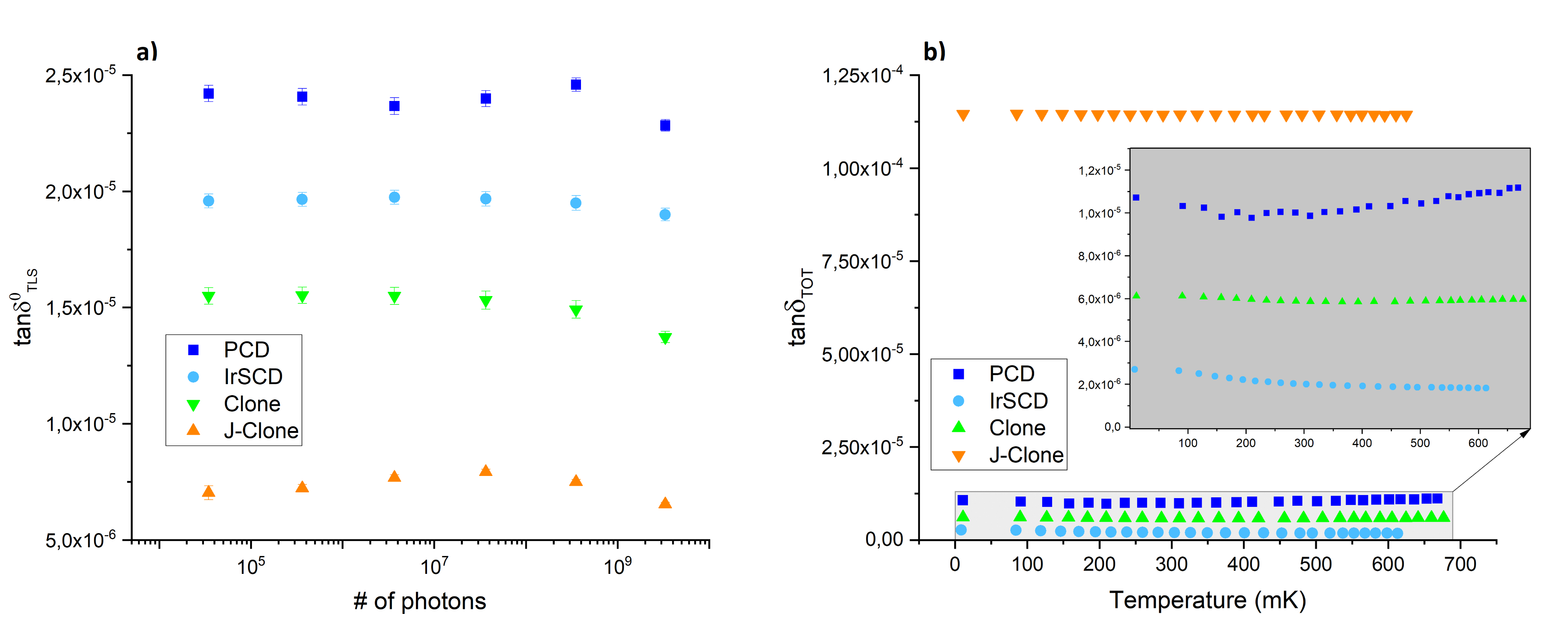} 
		\caption{a) values of $\tan \delta^0_{TLS}$ for the four samples, extrapolated from the fractional resonance frequency shift and b) lowest measured values of the total loss tangent $\tan \delta$}
		\label{pic3}
	\end{figure*}

	In fig. \ref{pic3}a  the values of $\tan \delta^0_{TLS}$ extracted from the fitting of eq. \ref{TLS} are reported. As expected, the PCD shows the highest losses due to the increased number of defects and grain boundaries inherent to its polycrystalline nature.
	The IrSCD is the next worst performer of the group, followed by the Clone crystal and finally by the J-Clone, the only sample to show TLS related losses in the order of $10^{-6}$. We attribute the losses observed in the single crystal samples primarily to nitrogen-related point defects incorporated during the growth and joining processes. Therefore, the measured $\tan\delta^0_{TLS}$ seems to have a direct link to their nitrogen inclusion levels. It´s worth to note that not all possible dielectric loss channels are associated with TLS dynamics. Fig. \ref{pic3}b demonstrates this: the values of the total loss tangent calculated with Eq.\ref{tan} paint a different picture, with the "J-Clone" sample being now the worst performer. This is somewhat expected, since we placed the resonator across the boundary between two tiles, where the defects density is larger and local stresses confined to 100-200 $\mu$m region \cite{Yamada2012} around the joint (absent in the "clone" sample) produce boundary-induced scattering that can dominate the total dielectric loss even when TLS contributions are minimized. The other samples are close in performances, and they range between 10$^{-5}$ of the PCD and roughly 2$\cdot 10^{-6}$ of the IrSCD.
	In Table \ref{tantable} we report the average values of both loss mechanisms for all the samples.

	\begin{table}[h]
		\caption{Average values of $\tan \delta^0_{TLS}$ and $\tan \delta$ for all the samples.}
		\begin{ruledtabular}
			\begin{tabular}{|c|cccc|}
				
				&	PCD &   IrSCD  & Clone   & J-Clone\\
				\hline
				$\tan \delta^0_{TLS}$ & $2.39 \cdot 10^{-5}$ & $1.9525 \cdot 10^{-5} $ & $ 1.508 \cdot 10^{-5} $ & $ 7.31 \cdot 10^{-6} $\\
				\hline
				$\tan \delta$ & $ 1.04 \cdot 10^{-5} $ & $ 2.04 \cdot 10^{-6} $ & $ 5.94 \cdot 10^{-6} $ & $ 1.44 \cdot 10^{-4} $ \\

			\end{tabular}
		\end{ruledtabular}
		\label{tantable}
	\end{table}
	
\begin{figure*}
		\includegraphics[width=\textwidth]{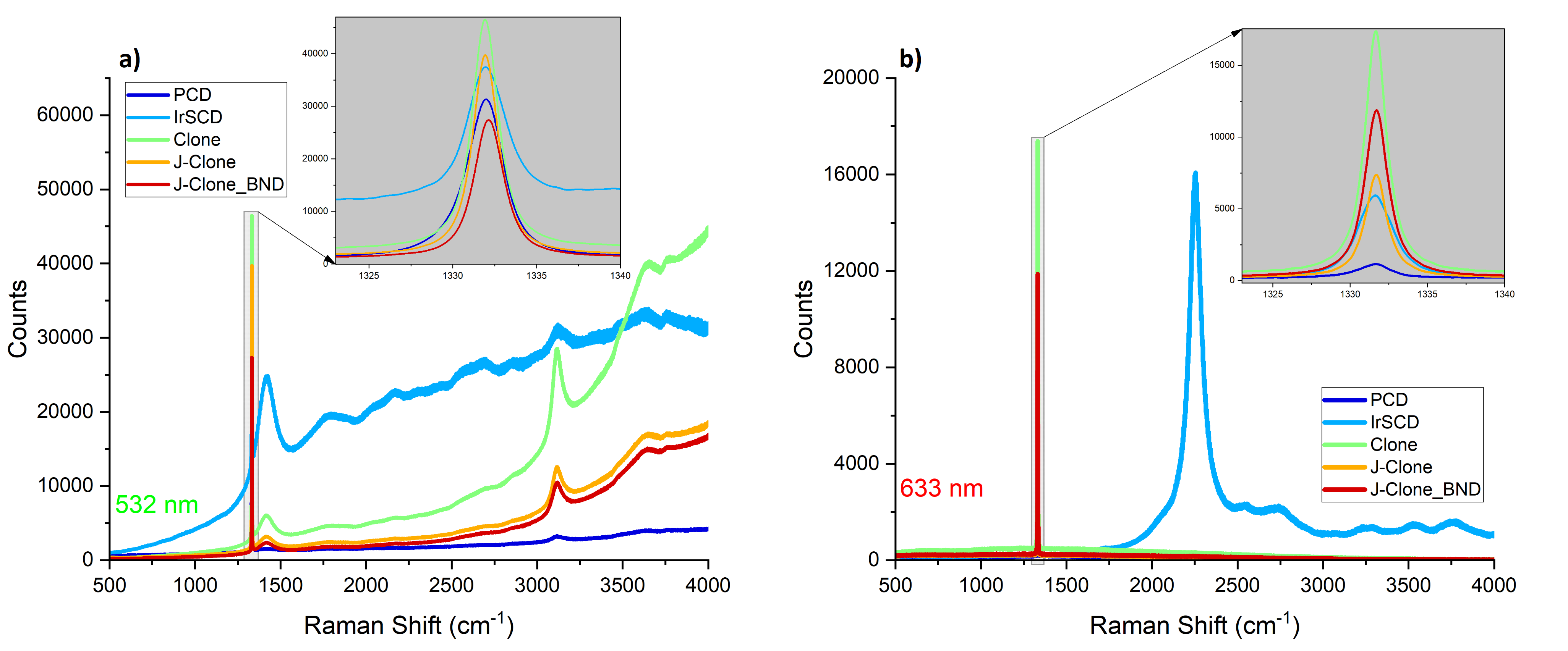} 
		\caption{Raman measurements of the diamond samples performed with a) green (532 nm) and b) red (633 nm) excitations. BND refers to measurements performed focusing the laser light directly on the boundary between two tiles of the J-Clone.}
		\label{pic4}
	\end{figure*}

	In light of these results, we performed multi-color (532 and 633 nm) Raman spectroscopy (fig. \ref{pic4}) to evaluate if we could associate the measured losses with the presence of impurities or defects in the samples. All samples show a very defined response at 1332 cm$^{-1}$ under both excitations as expected. This peak can be influenced by the presence of the $\mathrm{NV}^0$ color-center line, which lies so close to the first-order diamond line that the two structures end up overlapping \cite{Jiang2024}. The Clone sample shows under both lights the strongest signal, an indication of the high degree of crystallinity that a high quality HPHT seed can provide for the CVD growth.
	Under red light, the IrSCD sample exhibits a very large peak at around 2250 cm$^{-1}$, due to the C$\equiv$N stretching mode, associated with nitrile bonds\cite{Chowdhury1998}. This indicates that nitrogen is strongly interacting with carbon within the lattice resulting in a very stable form. These observations are in agreement with the resonance measurements that show that the IrSCD has the highest TLS losses among the single crystal samples. The smaller shoulders up to 2730 cm$^{-1}$ after the C$\equiv$N peak correspond to combination bands involving CN-related vibrations \cite{Chowdhury1998}, while the structures beyond 3000 cm$^{-1}$ are likely attributed to O-H or C-H stretching in surface hydroxyl or hydrocarbons groups \cite{Chowdhury1998, Yang2019}, sign of a certain degree of atmospheric surface contamination. 
	Finally, the structures in the 3500 - 3700 region are typically associated with sidebands of zero phonon line (ZPL), relative  $\mathrm{NV}^-$  centers \cite{Ren2024}.  These structures are absent in all other samples under 633 nm irradiation, which suggest that nitrogen is in these cases  incorporated less deeply and not as strongly. Under 532nm excitation, a new peak appears at 1420 cm$^{-1}$ for all samples. Generally associated to $\mathrm{NV}^+$ centers zero-phonon line \cite{Jiang2024}, the intensity of this peak signal follows the trend outlined by the resonators measurements, again with higher nitrogen content samples associated to higher TLS losses, with the exception of the PCD. Under green excitation we also observe a strong luminescence background typically associated with NV-centers activity, and whose intensities can be directly correlated to the losses measured with the resonators. None of the recorded spectra show any significant peak associated with graphite or carbon chains. While we expected this from the SCD substrates, it´s not the case for the PCD, which should show signs of non-diamond like carbon bonds. That said, if we consider the relatively large grain size (> 50 $\mu$m on the growth side) of the poly-crystalline sample, it is plausible that the Raman laser spot is fully contained within a single grain, which has virtually no sp2 bonds within it. Finally, in the region > 3000 cm$^{-1}$ we have the features deriving from C-H bond (peak at 3120 cm$^{-1}$) and O-H stretching vibrations from 3650-3750 cm$^{-1}$ \cite{Ferrari2000, Yang2019}. As a last consideration, the uniformity of the J-Clone sample (orange and red curves in Fig.\ref{pic4}a) is rather remarkable: the slightly lower signal coming from the boundary region can be explained in terms of refractive effects induced by the discontinuity that alter the local reflectivity of the sample but other than that, spectra taken above the two different areas basically coincide. 
	
	\section{Conclusions}\label{conc}

	In this work we presented loss characterization of synthetic diamond samples grown with different techniques via superconducting lumped-elements micro-resonators. The fitting model used to analyze the data is based on TLS theory and enables the determination of the TLS loss tangents of the target substrates by only measuring the frequency shift at ultra low temperatures. All of our single-crystalline samples show lower TLS related losses compared to the standard commercially available poly-crystalline reference. In terms of total loss tangent, the J-Clone shows higher values compared to the other samples due to the increased density of defects in the joint region. The other single crystal samples exhibit values in the order of 10$^{-6}$ as expected. Complementary Raman spectroscopy measurements allowed us to associate increased nitrogen contents with increased TLS losses.  In particular, the Raman spectrum of the IrSCD shows a very strong response at 2250 $\mathrm{cm}^{-1}$ which is associated to the triple C$\equiv$N bond, an indication of a very stable nitrogen inclusion during this particularly innovative growth process. The other two samples perform better in terms of TLS losses, and the absence of the 2250 cm-1 peak in the Raman spectra indicates that less strongly bonded, more superficial defects dominate these samples.
	
	\section{Acknowledgments}
	This work was partially funded by the Deutsche Forschungsgemeinschaft (DFG, German Research Foundation) – Projektnummer (project number) 467785074.
	
	\section*{Data Availability Statement}
	
	The data that support the findings of this study are available from the corresponding author upon request.
	

	\nocite{}
	\bibliography{Diam.Diel.Char.AIP}
	
\end{document}